# Design-Reality Gap Analysis of Health Information Systems Failure

Hanyani Makumbani[a], Pitso Tsibolane[a]*

[a]*Information Systems Department, Commerce Faculty, University of Cape Town, South Africa*

**Abstract**

This study investigates the factors contributing to the failure of Health Information Systems (HIS) in a public hospital in South Africa. While HIS have the potential to improve healthcare delivery by integrating services and enhancing effectiveness, failures can lead to service interruptions, revenue loss, data loss, administrative difficulties, and reputational damage. Using semi-structured interviews with key stakeholders, we employed a hybrid data analysis approach combining deductive analysis based on the Design-Reality Gap Model and inductive thematic analysis. Our findings highlight several factors contributing to HIS failures, including system capacity constraints, inadequate IT risk management, and critical skills gaps. Despite these challenges, end users perceive HIS positively and recommend its implementation for streamlining daily processes. This study underscores the importance of addressing design-reality gaps to improve HIS outcomes in public healthcare settings.

* Corresponding author. Tel.: +27 21 650 1542
  E-mail address: pitso.tsibolane@uct.ac.za

# 1. Introduction

Public healthcare institutions in South Africa continuously strive to transition from manual paper-based records management systems to Health Information Systems (HIS) and Electronic Health Records (EHRs) to improve efficiency and patient care [13]. However, transitioning from paper-based to electronic systems is significantly challenging, resulting in failed HIS implementations in healthcare facilities [29].

This case study focuses on a South African public hospital implementing a cutting-edge HIS to effectively manage the patient's medical records. However, the HIS implementation encountered significant challenges that led to considerable disruption in healthcare services and financial losses for the hospital. The primary research question underpinning this research is: "What factors contribute to the failure of Health Information Systems (HIS) in South African public hospitals."

The next section of the paper will cover the literature review, followed by a discussion of the adopted theoretical framework. The following section will present the findings, a discussion of the findings, and the study's conclusion.

# 2. Literature Review

## *2.1. Health Information Systems (HIS)*

HIS are digitized systems for managing, storing, and transmitting health-related information. They gather, analyze, and store patient medical and health outcome-related data [3, 4, 7]. HIS encompasses electronic health records (EHRs), medical decision-support systems (MDSS), telemedicine systems, health information exchanges (HIEs), and other instruments that aid in the management of patient care and population health by medical professionals. In addition to supporting quality enhancement, population health management, and research, HIS can provide analytical tools. These systems are crucial instruments for healthcare organizations and can enhance the treatment of patients, increase effectiveness, minimize errors in diagnosis, and minimize healthcare costs [4, 27]. Health Information Systems (HIS) are a major enabling factor in the health service delivery landscape. These systems, also known as health management information systems (HMIS), are used for processing and storing patient information. Health information systems are classified into EMR (electronic medical records) or EHR (electronic health records) [1, 11]

## *2.2. Benefits and Challenges of HIS Use*

HIS can transform the health care system from a predominantly paper-based sector to one that uses clinical and other information to assist medical professionals in delivering superior treatment to patients. HIS also allow medical professionals to view comprehensive and reliable patient data, allowing them to make informed choices regarding treatment. This may contribute to improved patient outcomes and superior treatment quality [3, 41]. Adopting HIS may optimize administration tasks such as financial planning, financial administration, invoicing, collaboration among medical professionals, and maintaining patient records [5, 6, 20]. It has been proven to reduce the effort and time required to execute these tasks manually [7]. HIS further assists with collecting, administering, and analyzing health information, allowing hospitals to recognize trends and patterns and make informed choices to enhance patient treatment and results [11, 20, 22].

HIS also presents challenges, such as data security and privacy concerns, interoperability issues, and the need for ongoing maintenance and enhancements. The HIS contains sensitive personal data and needs to be regulated. The Protection of Personal Information Act (POPIA) governs how organizations securely gather, store, and transmit personal information. With the introduction of POPIA, HIS needs to ensure adequate controls are in place to mitigate the leakage of patient information. Cyber-attacks continue to rise, and perpetrators are finding sophisticated means to attack organizational systems. Most HIS are also adopting artificial intelligence, and attacks on Artificial Intelligence systems continue to increase [10, 19]. The challenges have added complexity to the implementation and maintenance of HIS due to the number of skills required to develop and maintain a compliant system, and therefore, implementation costs have escalated [15, 40].

## 3. Theoretical Framework

Prior research indicates that this concept can be used in design to analyze and resolve discrepancies between the intended design and the actual experience or implementation of a product, system, or service [2, 16, 18, 25]. The chosen model (shown in Figure 1) is particularly well-suited for this study. It effectively demonstrates the criticality of post-adoption expectations regarding information systems and the dynamic nature of these expectations as users accrue greater experience with the specific information system in question [8, 18, 28]. However, it is currently understood that the failure of Health Information Systems (HIS) is critical [27]. The study [17], therefore, formulates and elaborates upon the theoretical constructs of the "Design-Reality Gap" framework.

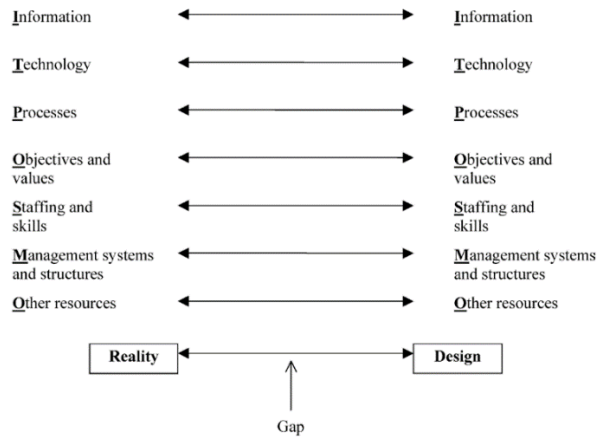

Figure 1: The design reality gaps [17, 25].

The framework consists of seven "ITPOSMO" dimensions (see Figure 1), which were used to investigate the factors contributing to the failure. The ITPOMSO model addresses seven critical dimensions to bridge design-reality gaps: Information (data management), Technology (broad ICT applications), Processes (user and stakeholder activities), Objectives and Values (stakeholder interests and cultural factors), Staffing and Skills (human competencies), Management Systems and Structures (organizational frameworks), and Other Resources (time and financial assets) [18, 25].

## 4. Results

A hybrid data analysis approach was employed, combining deductive analysis grounded on the Design Reality Gap Model and inductive thematic analysis of interview data. The study involved conducting semi-structured interviews with eight participants who were active users and played a crucial role in the failed HIS at the selected public hospital in South Africa. The relevant participant details are displayed in Table 1 below.

| Participant ID | Designation | Education | Experience at the hospital |
|---|---|---|---|
| RESP1 | Deputy Director: Information Technology | Advanced Diploma in Public Management<br>National Diploma in Information Technology<br>BTech: Incomplete | 17 years |
| RESP2 | Chief Admin Clerk | Diploma in Business Management | 15 years |
| RESP3 | Chief Admin Clerk | National Diploma in Financial Management | 9 years |
| RESP4 | Billing Clerk | Matric | 11 years |
| RESP5 | Maternity Clerk | Matric | 4 years |

| Participant ID | Designation | Education | Experience at the hospital |
|---|---|---|---|
| RESP6 | IT Systems Administrator | Diploma in Computer Engineering<br>BCom (In Progress) | 10 years |
| RESP7 | Cashier and Billing Clerk | Matric | 14 years |
| RESP8 | Operating Theatre Operations Manager (Nurse) | Diploma in Nursing | 12 years |

Table 1: Participants' background information.

### 4.1. Information Dimension

#### 4.1.1. Extent of data use

There are varying perspectives on the degree to which data was utilized. Notably, RESP1 highlighted that data usage was suboptimal and had the potential to yield greater value.

In explaining the extent of data use, RESP 2 & 3 mentioned that *"the data was used for various purposes, the data was mainly used for reporting (monthly), this includes reporting on patient admissions and discharges, classifications (Living Standards Measure), and budgeting."*

Meanwhile, RESP 1 said: *"The HIS data was supposed to be used for decision-making by management. Unfortunately, the data is not being used as an asset. The data is supposed to be analyzed and used to improve service delivery; unfortunately, this is not happening. Further, the data was supposed to be used to track patient admission stays and understand the cause of long admissions. It was supposed to be used for gap analysis, budgeting, and strategy planning."*

#### 4.1.2. Data quality

One of the main advantages of utilizing HIS is the ability to collect data and leverage it to perform analyses, enhance operations, and introduce efficiencies, thereby improving the quality of care that a healthcare provider can provide to patients. Participants who were patient-facing generally rated the data quality as low. The primary reason cited was the errors picked up by transcribing hand-written information onto the HIS as some data capturing was delayed. RESP1 noted that the data produced was *"90% accurate"*, citing that *"there were certain fields that were not mandatory could help track key elements about patients. Clerks are not working 24/7, and the hospital is open 24/7, with patients coming in anytime. The Clerks work from 7 am – 4 pm; when patients come to the hospital after the Clerk's knock-off time, Nurses complete the patient demographics on the book, and clerks are expected to capture the data in the HIS in the morning. The issue comes when the handwriting is not clear. This compromises the data quality. Until identity verification systems are implemented, the data accuracy will never be accurate."* Other participants noted that the data was generally accurate and aided decision-making.

### 4.2. Technology Dimension

#### 4.2.1. Frequency and use of HIS

However, some modules were underutilized, with the focus on managing patient data across the hospital. One respondent mentioned that he formed part of the IT department and was responsible for supporting the system. Another respondent, RESP 6, supported this, mentioning that *"the system was used for stock ordering, and that as a system administrator, they would use the system daily to reset user passwords."*

Most respondents echoed the prior statement that they used the system to manage patient data and billing. RESP4 said: *"I was using the system on a full-time basis for billing, allocation of patient funds…I was also using the system to open a new patient file, update patient information, complete patient bills, inclusion itemized billing. Most of the patients are expected to settle their medical consultation bill."*

RESP 2 & 3 mentioned that in their team as chief admin clerks, they had about 180 clerks allocated to each department and ward across the hospital. "*The clerks used the system on a daily basis for various purposes, including managing patient data and stock.*" RESP 8, who operated in the theatre environment, supported this view and used the system to manage medical equipment.

Lastly, RESP 7 mentioned that they "*used the system to receive cash from the patient and for billing purposes. As the hospital accepted both cash and cards*".

In summary, various participants leveraged the system to fulfill their roles. There is no consensus on the system's intended use. Most participants cited frequent use of the system, except for doctors who found it time-consuming.

*4.2.2. The extent of user requirements consideration in the design of the failed HIS*

Most participants were seemingly unaware of the requirements-gathering process undertaken prior to the implementation of the HIS. RESP1 & 2 cited that the system was already implemented when they started working at the hospital. This raises concerns about the transfer of Intellectual Property and lack of institutional memory across generations of users, which should have formed part of the implementation's scope.

Some respondents were aware of requirements gathering sessions and ongoing consultation, as noted by RESP4: "*There were consultations that took place on an annual basis. Updates and upgrades were also discussed by the consultant assigned by the third party to the hospital.*"

*4.2.3. Vulnerability scanning*

Scanning vulnerabilities is another function without alignment in the IT department. Both RESP1 and RESP6 give different versions of the truth. RESP1 said, "*Vulnerability scanning is outsourced to the e-government department. The e-government department oversees developing and maintaining e-government infrastructure, such as data centers, networks, and communication systems.*" Meanwhile, RESP6 mentioned, "*There was a system used to monitor the server. We used both Symantec to Microsoft Defender*".

*4.3. Processes dimension*

*4.3.1. Diagnosing the failure*

The researcher found that the system had reached its end of life and was no longer supported by the manufacturer. The system's failure may be attributed to the compromise it encountered, which was a direct consequence of the outdated infrastructure. The participants exhibited contrasting views of the signs that were provided by the system prior to its failure. Certain participants believed that the system saw a notable decrease in speed, but others asserted that the system exhibited no indications of malfunction.

RESP1 mentioned: "*There was a sequence of events that ultimately led to the system failure. There were signs on several occasions. At some point, the server could just stop working out of the blue, and no one understood Linux at the hospital. The failed HIS was Linux-based and there were no skills to maintain the system after the contract with the service provider expired. The data retrieved was only from 01$^{st}$ April 2017 until 11$^{th}$ November 2022. Patient data dating back to system implementation in 1999 has been lost.* "

According to RESP2 & 3: "*There were no signs that the system would crash. We came in on the Monday and the system was not functioning anymore.*"

*4.3.2. Impact of the system failure on processes and productivity*

The failure of the system resulted in various losses for the hospital including additional time, developing manual work arounds, financial losses and data losses that presented a reputational risk to the hospital. According to RESP1: "*The hospital had to adopt manual workarounds whereby Excel documents were used to manage patient information. A lot of losses were encountered because of the failure, including financial, reputation, etc.*"

RESP2 & 3 mentioned that "*...manual workarounds had to be adopted and downtime manual registers were used to register patients coming to the hospital. A total of 177,000 patients had to be manually captured in the new system. 80% of the patients have now been captured.*"

System enhancements were believed to be outsourced to the supplier; however, one participant had a negative view of this approach as they believed that the government should empower internal resources to complete this function.

*4.3.3. Specific tasks that were difficult or cumbersome on the failed HIS*

Participants had varying views, with a major deficiency noted around the consolidation of patient data by category. RESP4 was very pleased with the system and said: *"The failed system was extremely user friendly; the newly implemented HIS system is more complex and below user friendly."* RESP 2 & 3 were not entirely happy and mentioned, "*The challenge with the failed HIS is that it doesn't do any consolidation per classification. In essence, it didn't consolidate patients in the same category; for example, it doesn't group cross-border patients, external funders, correctional service patients, etc. The system doesn't segment patients. This would have to be done manually and took a long time."*

*4.3.4. Failures in backup systems that should have prevented the failure.*

Participants believed reliance on the central function and key man dependency resulted in a lack of accountability and oversight by the internal teams, resulting in a gap in backup processes. RESP1: *"The backup of the HIS was centralized to the provincial office in Johannesburg. The backup was not done on-prem at the hospital. The system was Linux based; the hospital has no skills to manage backups on the Linux environment."*

*4.4. Staffing and Skills*

*4.4.1. Lack of advanced technical expertise*

Respondents indicated a lack of Linux skills in managing backups inside the hospital and stated the presence of a physical backup server within the hospital. The lack of clarity on the management and ownership of backups is concerning. It can be assumed that the team managing the failed HIS was not clear on these matters.

According to RESP1: *"The backup of the HIS was centralized to the provincial office in Johannesburg. The backup was not done on-prem at the hospital. The system was Linux based; the hospital has no skills to manage backups on the Linux environment."*

*4.5. Objectives and values*

*4.5.1. The objective of the system and if it was achieved.*

RESP2 and RESP3 said: *"The system served the purpose it was meant for. However, there are capabilities that were not available. Some of these include patient identity verification."*

*4.6. Management systems and structures*

*4.6.1. Accountability for system availability*

Participants believed that a lack of oversight by the internal team contributed to the failure of the HIS due to the system being managed by a central team. RESP1: *"We have tools to monitor the systems under our responsibility. We use a tool called PRTG to manage servers and switches. These systems are under our management. These stats are presented in the MANCO. Unfortunately, the failed HIS was managed by the central team based in Johannesburg."*

*4.6.2. Key risk indicators to ensure adequate operation of systems.*

The participant indicated the absence of risk indicators to manage IT risk. RESP1: *"There are no key indicators to manage IT risks in the hospital."*

*4.7. Other resources*

*4.7.1. Budget set aside for system development and maintenance*

Participant believed that there was no budget, or insufficient resourcing allocated to system development and maintenance.

RESP1: *"Maintenance of system purely sat with the 3rd party as well as the central function managing HIS in the provincial government in Johannesburg."*

## 5. Summary of Findings

The findings can be summarized as shown below:

| Dimension | Description | **Rating of the Design-Reality Gap** |
|---|---|---|
| *I*nformation dimension | Identifies the data being utilized by the HIS, in comparison to the information requirements included in the system's design with the information actually being utilized by the organization [18, 25]. | High |
| *T*echnology dimension | Identifies the hospital's technology based on a comparison between the HIS design requirements and the real situation [16, 25]. | High |
| *P*rocess dimension | Describes the hospital's work processes by comparing the processes required for the successful implementation of the HIS with the real situation [2, 25]. | Medium |
| *O*bjectives dimension | Compares the objectives and values required by stakeholders for the effective functioning of the HIS to their actual objectives and values [2, 25]. | Low |
| *S*taffing and Skill dimension | This dimension aims to establish the availability of necessary skills to maintain the HIS while comparing the skills to maintain HIS in the real situation [2, 18, 25]. | High |
| *M*anagement and Structure dimension | This dimension pertains to the necessary management systems and structures that are required, it involves a comparative analysis of the requirements for the effective management/maintenance of HIS versus the real situation [18, 25]. | Low |
| *O*ther resources | This dimension pertains to the assessment of the feasibility of the implementation and operation of a new application in relation to the current availability of resources, both in terms of time and financial considerations [18, 25]. This analysis considers financial, time, technological, and other limitations to assess practical implementation. [18, 25]. After the identification of prioritized gaps, the next step involves the development of solutions aimed at bridging these gaps. Potential resolutions may entail the modification of the design concept or exploration of substitute technologies [2, 18, 25]. | Low |

Table 2: Design Reality Gap dimensions.

## 6. Conclusion

This study looked at the factors that contribute to HIS failure in the South African context from the perspective of the Design Reality Gap framework. The data was collected from a South African hospital using semi-structured interviews and several findings were made. Respondents indicated that without policy change to address amongst others, IT risks and recruiting skilled resources, HIS failures will increase, furthermore, systems capacity constraints were a significant contributing factor to failure of HIS.

This research is likely to offer various potential contributions to theory in the field of Information systems specifically in the topic of health information systems.

The study had several limitations, one being that the researcher only had less than six months to collect data and also, the data was collected from a few participants in a South African hospital. Recommendations are for more research on innovations such as the use of artificial intelligence, blockchain technologies, and industrial control systems in HIS and the adoption of IT risk practices in government hospitals.